%% file: main.tex
\RequirePackage{docswitch}
\def\flag{lsstdescnote} %\input{``main.tex''}
\setjournal{\flag}
\documentclass[\docopts]{\docclass} %#lsstdescnote}

\usepackage{lsstdesc_macros}

\usepackage{graphicx}
\graphicspath{{./}{./figures/}}
\input{plasticc_macros}

\begin{document}

\title{The Photometric LSST Astronomical Time-series Classification Challenge ({\plasticc}): Data set}

\maketitlepre

\begin{abstract}
\input{abstract}
\end{abstract}

% Keywords are ignored in the LSST DESC Note style:
\dockeys{}

\maketitlepost

% ----------------------------------------------------------------------
% {% raw %}

% \section{Introduction}
% \label{sec:intro}
\input{introstuff}

\input{data_description}
% ----------------------------------------------------------------------

% \section{Methods}
% \label{sec:methods}

% ----------------------------------------------------------------------

%\section{Results}
%\label{sec:results}

% ----------------------------------------------------------------------

%\section{Discussion}
%\label{sec:discussion}

% ----------------------------------------------------------------------

%\section{Conclusion}
%\label{sec:conclusion}

% ----------------------------------------------------------------------

\subsection*{Acknowledgments}

%%% Here is where you should add your specific acknowledgments, remembering that some standard thanks will be added via the \code{desc-tex/ack/*.tex} and \code{contributions.tex} files.
\input{plasticc_acknowledgements}
%This paper has undergone internal review in the LSST Dark Energy Science Collaboration. % REQUIRED if true

% This work used TBD kindly provided by Not-A-DESC Member and benefitted from comments by Another Non-DESC person.

% Standard papers only: A.B.C. acknowledges support from grant 1234 from ...
\input{standard} % also available: key standard_short
\input{contributions} % Standard papers only: author contribution statements. For examples, see http://blogs.nature.com/nautilus/2007/11/post_12.html
% This work used some telescope which is operated/funded by some agency or consortium or foundation ...

% We acknowledge the use of An-External-Tool-like-NED-or-ADS.

%{\it Facilities:} \facility{LSST}

% Include both collaboration papers and external citations:
%\bibliography{main,lsstdesc}

\end{document}

%% file: plasticc_macros.tex
\newcommand{\plasticc}{\tt{PLAsTiCC}}
\newcommand{\numClasses}{14}
\newcommand{\numTotalClasses}{15}
\newcommand{\numObjectsTraining}{8000}
\newcommand{\numObjectsTest}{$\approx 3.5$~million~}
\newcommand{\specz}{\tt{hostgal\_specz}}
\newcommand{\hostphotoz}{\tt{hostgal\_photoz}}
\newcommand{\hostphotozerr}{\tt{hostgal\_photoz\_err}}
\newcommand{\class}{\tt{target}}
\newcommand{\passband}{\tt{passband}}

\newcommand{\distmod}{\tt{distmod}}

%% file: abstract.tex
The Photometric LSST Astronomical Time Series Classification Challenge
({\plasticc}) is an open data challenge to classify simulated astronomical
time-series data in preparation for observations from the Large Synoptic
Survey Telescope (LSST), which will achieve first light in 2019 and commence
its 10-year main survey in 2022. LSST will revolutionize our understanding of
the changing sky, discovering and measuring millions of time-varying objects.
In this challenge, we pose the question: \textit{how well can we classify objects in the sky that vary in brightness from simulated LSST time-series data, with all its challenges of non-representativity?} In this note we explain the need for a data challenge to help classify such astronomical sources and describe the {\plasticc} data set and Kaggle data challenge, noting that while the references are provided for context, they are not needed to participate in the challenge.

%% file: introstuff.tex
%\documentclass{article}
%\usepackage{graphicx}
%\usepackage[a4paper, total={7in, 9in}]{geometry}
%\usepackage{soul}
%\input{plasticc_macros}

%\begin{document}
\section{Introduction}

\label{sec:intro}
{\plasticc} is a large data challenge for which participants are asked to \textit{classify astronomical time series data}. These simulated time series, or `light curves', are measurements of an object's brightness as a function of time - by measuring the photon flux in six different astronomical filters (commonly referred to as passbands). These passbands include ultra-violet, optical and infrared regions of the light spectrum. There are many different types of astronomical objects (that are driven by different physical processes) that we separate into astronomical classes. The challenge is to analyze each set of light curves ($1$ light curve per passband, $6$ passbands per object) and determine a probability that each object belongs to each of these classes. The time-series data provided are simulations of what we expect from the upcoming Large Synoptic Survey Telescope (LSST\footnote{\href{https://www.lsst.org/}{https://www.lsst.org/}}). LSST is under construction high in the deserts of northern Chile on a mountain called Cerro Pachon. When complete, LSST will use an $8$~meter telescope, with a $3$ billion pixel camera to image the entire Southern sky roughly every few nights and over a ten-year duration. In order to prepare for the data onslaught, hundreds of scientists are joining forces to form collaborations and working groups including the Transient and Variable Stars Collaboration (TVS\footnote{\href{https://lsst-tvssc.github.io/}{https://lsst-tvssc.github.io/}}) and the Dark Energy Science Collaboration (DESC\footnote{\href{http://lsstdesc.org/}{http://lsstdesc.org/}}). As part of the collaborative effort, these two collaborations developed the {\plasticc} team\footnote{In particular, the science requirements of the LSST Project is given \href{https://www.lsst.org/scientists/publications/science-requirements-document}{online}, while the DESC Science requirements document is available \href{https://arxiv.org/abs/1809.01669}{on the arXiv}. The overview of LSST Science is summarized in the \href{https://arxiv.org/abs/0912.0201}{LSST Science Book.}}.

%Each night, an LSST data alert stream will be made available through the LSST data centers that will contain thousands of alerts flagging celestial sources that have changed brightness. 

The time-series data in this challenge are called light curves. These light curves are the result of difference imaging: two images are taken of the same region on different nights, and the images are subtracted from each other. This differencing procedure catches both moving objects (asteroids) that are discarded for this challenge, and also objects that stay at the same position but vary in brightness. Depending on when the object first exploded or brightened, the flux may be increasing or decreasing with time. %In addition, one can model the patch of sky including the galaxy (known as `scene modeling' photometry) to provide a reference to compare to the new image containing the source. 
%In general while flux should be positive relative to some quiesent phase, very faint objects may appear as negative fluxes (this is discussed more in the context of the {\plasticc} data files). 
The specific manner in which the flux changes (the length of time over which it brightens, the way the object brightens in different passbands, the time to fade, etc.) is a good indicator of the underlying type of the object. We use these light curves to classify the variable sources from LSST. The data are classified using a set of training data (light curves) for which the true classifications (i.e. labels) are given. The participants in {\plasticc} are asked to separate the data into {\numTotalClasses} classes, {\numClasses} of which are represented in the training sample. The models used to generate these classes will be described in an upcoming paper that will be released once the challenge is complete. The final class is meant to capture interesting objects that are hypothesized to exist but have never been observed and are thus not in the training set. This class would encompass any objects not seen before (i.e. there may be more than one type of object in this class).

Once their classification algorithms are tuned on the training set, the participants will apply those same algorithms to previously unseen `test' light curve data. The goal is to classify those objects into classes.

\begin{figure*}[htbp!]
\begin{center}
%\begin{tabular}{ll}
%\includegraphics[scale=0.4, trim = 15mm 45mm 10mm 20mm, clip]{figures/lcplot_model01a.pdf}
\includegraphics[scale=0.45]{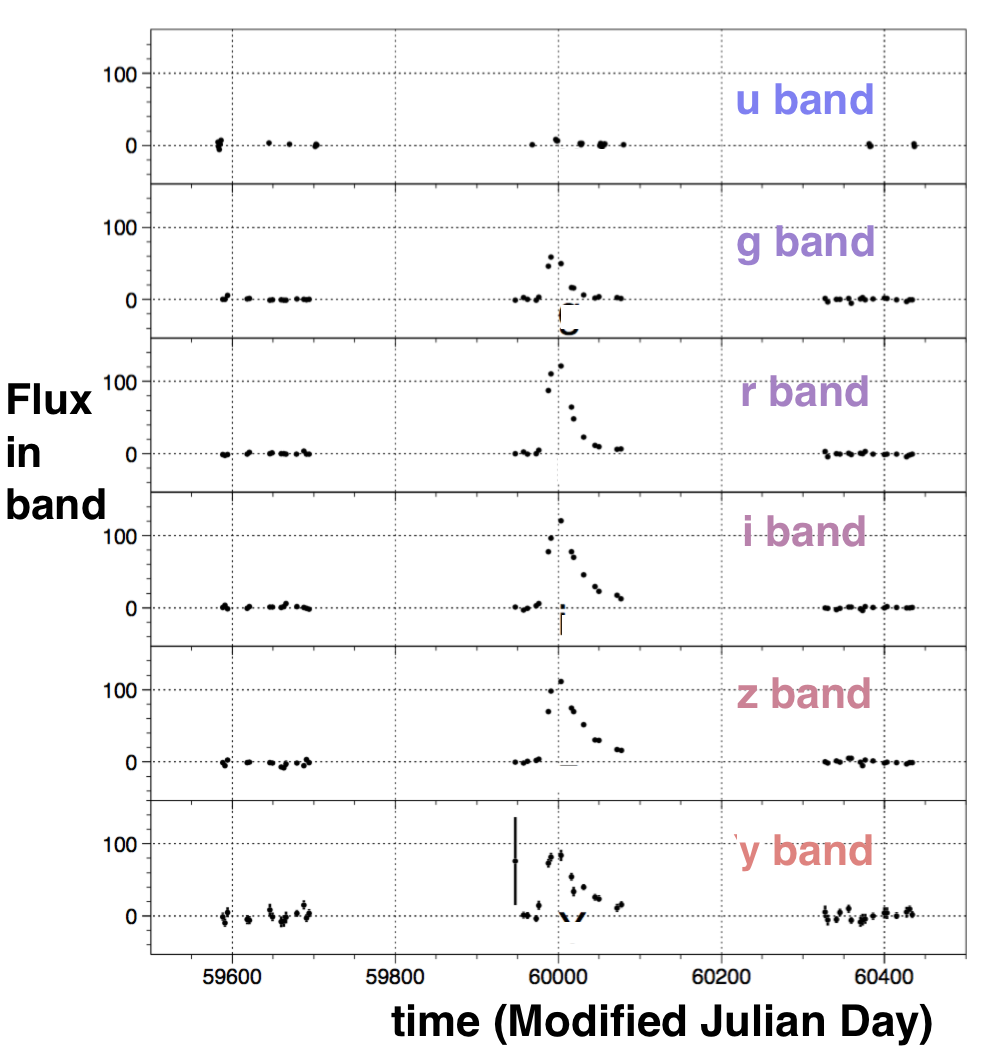}
\includegraphics[scale=0.45]{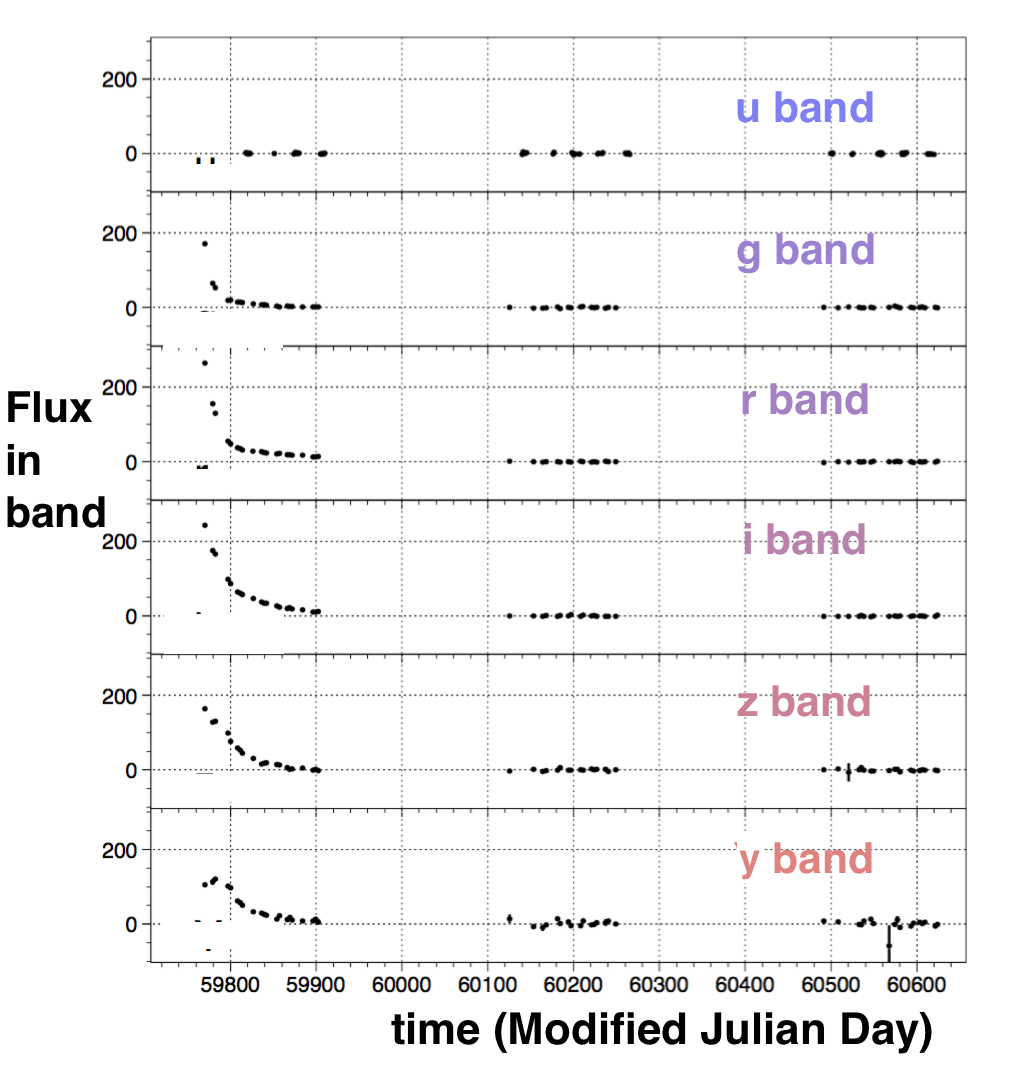} \\
\includegraphics[scale=0.45]{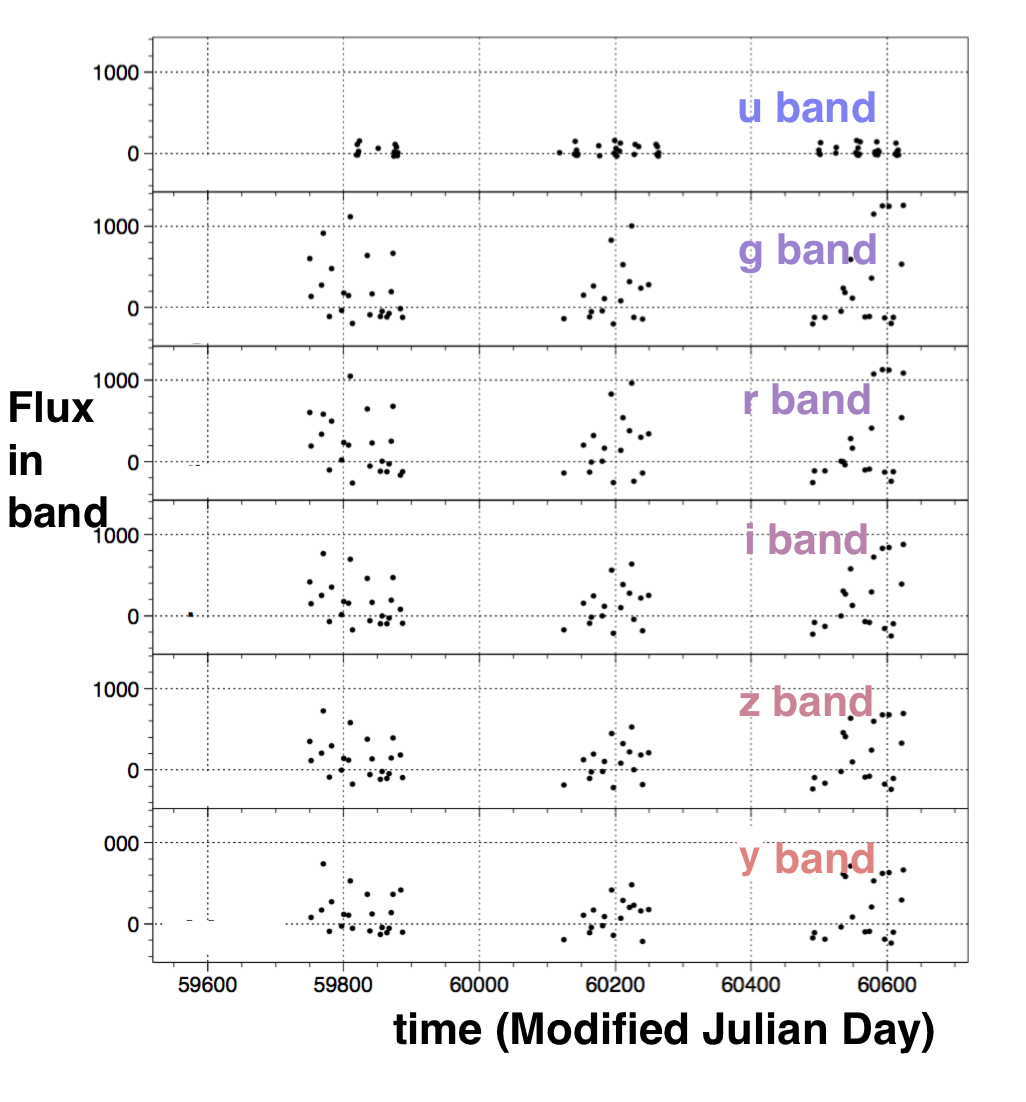} 
%\end{tabular}
\caption{Example light curves in the {\plasticc} data set. The three example objects display different changes in flux with time that are typical of real objects. The top row shows the same type of object that brightens from some `nominal' brightness and then fades away. The difference between the panels in the top row show the effects of observation gaps (linked to the telescope survey strategy): the top-right panel illustrates that the brightening of the flux can occur near observation gaps, and therefore may not include the full time period of brightening (or fading) for the object. 
The bottom figure illustrates an object that brightens and fades periodically rather than as a `once-off' event. In addition, all three panels show that seasonal gaps over a roughly two-year period, and the cadence of observations as set by the LSST survey strategy can introduce gaps in the light curve.\label{fig:lc}}
\end{center}
\end{figure*}

%For each object, the data include summary information: its position on the sky, an estimate of its observed redshift which correlates with its distance away from Earth, and other properties of the sky near the object. In addition, the light curve \textit{photometry} data on the object is a table of fluxes at different times of observation, and at different wavebands (i.e. the average energy of the light within a range of wavelengths). We go into more detail in the following section about the astronomical terminology used here.

In Figure~\ref{fig:lc}, we show three example light curves from the training set. The top two panels show sources that brighten and fade over a short time period. 
The lower panel shows a variable object that can fade temporarily, but always brightens again. It is worth noting that not all variables brighten and fade, some could just brighten and fade, never to rise again. Also note the gaps between observations: small gaps (from minutes to days or weeks) from the time between telescope observing a given patch of sky, and large gaps ($>6$ months) where the object is not visible at night from the LSST site. 

LSST will have two different kinds of survey regions: the so-called Deep Drilling Fields (DDF) are small patches of the sky that will be sampled often to achieve great depth (i.e. to be able to measure the flux from fainter objects). Objects in these DDF patches will have light-curve points that are extremely well determined and therefore have small errors in flux. The Wide-Fast-Deep (WFD) survey covers a larger part of the sky (almost $400$ times the area of the deep fields) that will be observed less frequently (and so light-curve points will have larger uncertainties), but will discover many more objects over the larger area.

\clearpage
\section{Astronomy background}
While we think of the night sky and the distant stars it contains as static, the sky is filled with sources of light that vary in brightness on timescales from seconds and minutes to months and years. 

Some of these events are called \textit{transients}, and are the
observational consequences of a large variety of astronomical phenomena. 
For example, the cataclysmic event that occurs when a star explodes generates a 
bright `supernova' signal that fades with time, but does not repeat. Other events are called \textit{variables}, since they vary repeatedly in brightness either in a periodic way, or episodically variable objects including active galactic nuclei (AGN) at the hearts of galaxies, pulsating stars known as Cepheids, and eclipsing binary stars that alternate blocking out each other's light from view.

%some other unknown brightness variation. 
% and originate from physical process governing high density regions of the Universe such 

These variation in these bright sources can provide important clues about themselves and their environment - as well as the evolution of the universe as a whole. For example, measurements of type Ia supernovae light curves provided the first evidence of accelerated expansion of the Universe. Each type of transient and variable provides a different clue that helps us study how stars evolve, the physics of stellar explosions, the chemical enrichment of the cosmos, and the accelerating expansion of the universe.  Therefore, the proper classification of sources is a crucial task in observational astronomy - especially in light of the large data volumes expected for the next generation of astronomical surveys - that includes LSST.

The question we address in this challenge is: \textit{how well can we classify astronomical transients and variables from a simulated light curve data set designed to mimic the data from LSST?} Crucially, the classifications will occur on a large test set, but the training data will be a small subset of the full data, and will also be a poor representation of the test set, to mimic the challenges we face observationally.

\subsection{Different methods for observing astronomical objects}
\label{subsec:observmethods}
Here we give more detail on LSST, and the challenge at hand. Two important modes for characterizing light from astronomical objects are called `spectroscopy' and `photometry.'

Spectroscopy measures the flux as a function of wavelength and is the modern equivalent of using a prism to separate a beam of light into a rainbow of colours. It is a high-precision measurement that allows us to identify emission \& absorption features indicative of specific chemical elements present in an object.  Spectroscopy is also the most accurate and reliable tool that enables classification of astronomical transients and variables. However being paramount for the classification task, spectroscopy is an extremely time-consuming process - with exposure times that are much longer than needed to discover objects with filters on an equivalently sized telescope. 
%photometrically 

Given the volume of data expected from the upcoming large-scale sky surveys, obtaining spectroscopic observations for every object is not feasible. An alternative approach is to take an image of the object through different filters (also known as passbands), where each passband selects light within a specific (broad) wavelength range. This approach is called photometry, and data obtained through this method are called photometric data. % (e.g. a redshift that is determined through photometry alone is called a photometric redshift).
 
\begin{figure}
\includegraphics[width=\textwidth]{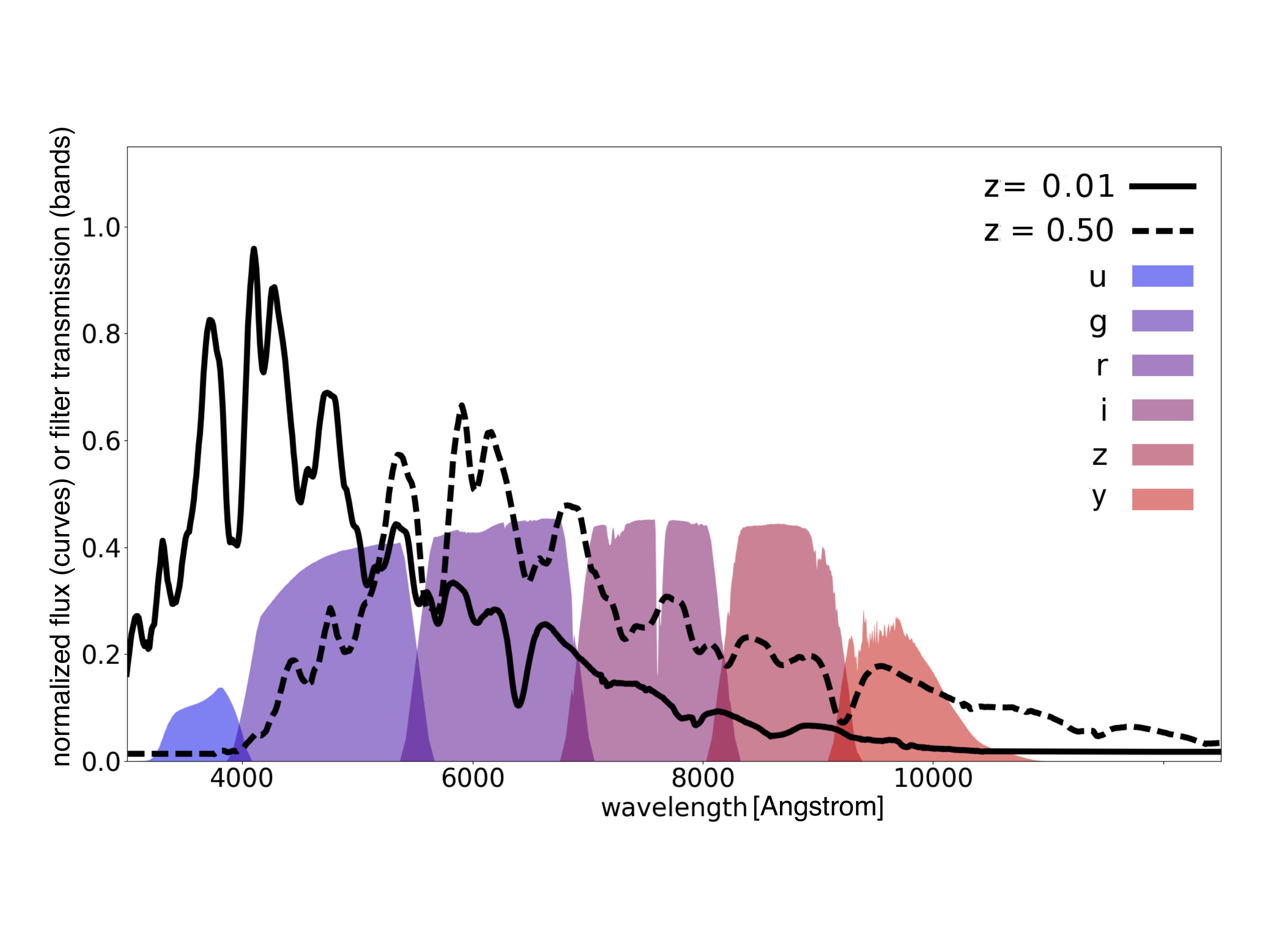}
\caption{Shaded regions show $ugrizy$ filter transmission (the fraction of light that is transmitted by the filter) vs. wavelength.
Each measured filter flux is a sum of the photons collected within the wavelength range.
Black curve shows a spectrum for a nearby Type Ia supernova at redshift $0.01$, while the dashed curve shows a spectrum for the same object at a larger redshift $0.5$: the spectrum of the object at higher redshift is shifted to longer wavelengths and the spectrum is also stretched compared to the low-redshift object (the galaxy at high redshift appears `redder').
The dashed spectrum {brightness is much lower than that of the solid spectrum, and has been scaled} to show its shape relative to the solid spectrum.}
\label{fig:filters}
\end{figure}

For LSST there are six passbands denoted $u,g,r,i,z,y$, that select light within different wavelength ranges: wavelengths between $300$ and $400$ nanometers for the $u$ band, between 
$400$ and $600$ nm for the $g$ passband, between $500$ and $700$ nm for the $r$
band, between $650$ and $850$ for the $i$ band, between $800$ and $950$ nm for 
the $z$ band, and between $950$ and $1050$ nm for the $y$ band.
The filter efficiencies vs. wavelength are shown in Fig.~\ref{fig:filters}.
For reference, the human eye is sensitive to light in the $g$ and $r$ bands.
The flux of light in each passband, measured as a function of time, is a light curve.
Classification is performed on these light curves. 
While a spectroscopic measurement can have great detail, passband data are one brightness measurement per band per exposure.
The challenge is to classify objects with the low (spectral) resolution light-curve data. 
Compared with spectroscopy, the advantage of measuring light curves with passband observations is that 
we can observe objects that are much further away and much fainter, and that one can observe many objects in a larger field of view at the same time (rather than measuring a spectrum of one or a few objects at a time). 

%In addition,
%light curves from LSST can be measured over half the sky, a much larger region than spectroscopy can cover.

% Photometry records how bright the source is at a given moment. The photometric information is encoded as the flux (energy from the object). The photometric light curve has six pieces of information, namely the flux in six wavelength bands (named $ugrizY$) at any moment in time.
% These photometric wavelength band fluxes are the integrals of the spectrum over the filter bandpasses of atmosphere and of the instrument divided by the energy of photons in the central wavelength of the filter. A sequence of photometric observations made at different times is called a light curve. It measures how the energy of the source evolves with time and can also be used to characterize different types of astronomical transients. As a consequence, for each object we will have a number of light curves in each filter (or band). Wavelengths are measured in units of Angstrom ($\AA$), where $1\AA = 10^{-10} m.$ Each band corresponds to a `color', with a width of around 1000 $\AA$, with the full set ranging from $3000\AA$ (blue light) to $9000\AA$ (near infrared light).

Beware that observations are sometimes degraded by moonlight, twilight, clouds, and atmospheric effects (that we call `seeing').
These degradations result in larger flux uncertainties, 
and this information is included in the light-curve data (see Sec.~\ref{sec:thedata} below).
In the challenge data sets, in addition to light curves, two other pieces of information are provided for each object.

%In addition to providing light curves, two other pieces of information are provided
%for each object. 

The first is called `redshift' (sometimes the symbol $z$ is used, that should not be confused with the passband in $ugrizy$). The cosmological redshift also slows down the rate of arrival of photons
compared to the rate at which they were emitted (called cosmological
time-dilation). An object at $z=1$ is $1.8$ times longer-lived than an object at $z=0.1.$ Because higher redshift objects are more distant, the
light we receive from them is fainter (and redder), and so we may actually see fewer observations above the signal-to-noise floor.  %The name comes from the fact that distant objects in the universe are redder due to the expansion of the universe; more distant objects have a larger redshift.

The training data include accurate redshifts for the objects, but the test-data redshifts
are approximate measurements based on $ugrizy$ passband measurements from the galaxy that contains the object (a galaxy that is located at the same position on the sky, and at the same redshift as the object).
While sources brighten and fade, the host galaxy fluxes don't change
and can thus be measured before a variable object it contains gets bright enough to be detected.

Beware that a few percent of the test data redshifts are catastrophic, 
meaning that some redshift uncertainties greatly underestimate the difference
between measured and true redshift.

The effect of redshift on light from a galaxy reaching earth is illustrated in Fig.~\ref{fig:filters}.
The black curve shows a {\it nearby} Type Ia supernova spectrum at a redshift of $0.01$,
corresponding to a distance of $140$ million light years. 
While the term `nearby' may seem strange in this case, this distance is 
indeed nearby when compared with the whole range of cosmic distances.
Visual inspection of the supernova spectrum and the filter efficiencies shows that
the maximum flux (spectrum summed over filter) is in the 
green ($g$) filter.
The dashed curve shows a spectrum from a more distant supernova, 
corresponding to a redshift of $0.5$, or $5.1$ billion light years 
away.\footnote{The relation between distance and redshift is not linear, 
but is a function derived from General Relativity that depends on the properties of 
dark matter and dark energy.}  % end footnote
The maximum flux is now shifted to the red ($r$) filter.
As the redshift and distance increase, the maximum flux appears
in a redder filter: hence the term `redshift.'

The second piece of additional information is related to extinction from our Galaxy, 
known as the Milky Way. Our light curve measurements are corrected for the atmosphere and telescope transmission (assuming that the telescope is calibrated off standard stars, knowledge of this isn't needed to participate in {\plasticc}). In addition, we correct each light curve for the absorption of light traveling through Milky Way `dust' on its way to Earth. This absorption is strongest in the ultra-violet $u$-filter, and weakest in the infrared filters ($izy$). We include the value of the Milky Way extinction at the sky coordinates of each object in the data release, with the label `MWEBV', that is an astronomical measure of how much redder an object appears compared to a Milky Way without dust. 
Larger MWEBV values correspond to more Milky Way dust along the line of sight to the objects, making the objects appearing redder. All objects in the {\plasticc} data are selected to have $\mathrm{MWEBV} < 3,$ to ensure that we are not looking too close to the disc of the Milky Way, or similarly to ensure we are not looking through a large amount of dust\footnote{A value $\mathrm{MWEBV} = 3$ means that a large percentage of the light from an object ($\simeq 99\%$ of the light depending on the spectrum of the object) is absorbed by the dust; only the brightest objects could be seen through such high levels of dust}.

%% file: data_description.tex
\section{The data}
\label{sec:thedata}
The {\plasticc} data are separated into a training data set and a test set; the latter is the data without classifications that needs to be classified. The data are provided in multiple CSV files, that are accessed from the Kaggle website.\footnote{The URL for the competition is \href{https://www.kaggle.com/c/PLAsTiCC-2018/}{https://www.kaggle.com/c/PLAsTiCC-2018/}} There are two types of files:
\begin{enumerate}
\item header files that contain summary (astronomical) information about the objects
\item light-curve data for each object consisting of a time series of fluxes in six filters ($ugrizy$), including flux uncertainties
\end{enumerate}

The header file lists each source in the data indexed by a unique identifier 'objid', that is an integer. Each row of the table lists the properties of the source as follows: 
\begin{itemize}
\item {\tt{object\_id}}: the Object ID, unique identifier (given as int32 numbers).
\item {\tt{ra}}: right ascension, sky coordinate: longitude, units are degrees (given as float32 numbers).
\item {\tt{decl}}: declination, sky coordinate: latitude, units are degrees (given as float32 numbers).
\item {\tt{gal\_l}}: Galactic longitude, units are degrees (given as float32 numbers).
\item {\tt{gal\_b}}: Galactic lattitude, units are degrees (given as float32 numbers).
\item {\tt{ddf}}: A Boolean flag to identify the object as coming from the DDF survey area (with value {\tt{ddf} $= 1$} for the DDF). Note that while the DDF fields are contained within the full WFD survey area, the DDF fields have significantly smaller uncertainties, given that the data are provided as additions of all observations in a given night.
\item {\specz}: the spectroscopic redshift of the source\footnote{In the heliocentric frame and including a contribution from the peculiar velocity of the host. We note again that in-depth knowledge of these details is not required for {\plasticc} participation.}. This is an extremely accurate measure of redshift, provided for the training set and a small fraction of the test set (given as float32 numbers).
\item {\hostphotoz}: The photometric redshift of the host galaxy of the astronomical source. While this is meant to be a proxy for {\specz}, there can be large differences between the two and {\hostphotoz} should be regarded as a far less accurate version of {\specz}. The {\hostphotoz} is given as float32 numbers.
\item {\hostphotozerr}: The uncertainty on the {\hostphotoz} based on LSST survey projections, given as float32 numbers.
\item{\distmod}: The distance (modulus) calculated from the {\hostphotoz} since this redshift is given for all objects (given as float32 numbers). Computing the distance modulus requires knowledge of General Relativity, and assumed values of the dark energy and dark matter content of the Universe, as mentioned in the introduction section.
\item {\tt{MWEBV = MW E(B-V)}}: this `extinction' of light is a property of the Milky Way (MW) dust along the line of sight to the astronomical source, and is thus a function of the sky coordinates of the source {\tt ra, decl}. This is used to determine a {\passband} dependent dimming and reddening of light from astronomical sources as described in subsection~\ref{subsec:observmethods}, and is given as float32 numbers.%, and based on the Schlafly \textit{et al.} (2014) dust model.
\item {\class}: The class of the astronomical source. This is provided in the training data. Correctly determining the target (correctly assigning classification probabilities to the objects) is the goal of the classification challenge for the test data. The {\class} is given as int8 numbers.
\end{itemize}

The second table of time-series data contains information about the sources and their brightness in different passbands as a function of time i.e. it is the light-curve data. Each row of this table corresponds to an observation of the source at a particular time and passband. There is one light-curve file for the training data, but $11$ light-curve files for the (much larger) test set. The test tables contain a mix of DDF and WFD objects ($1\times$ DDF and $10\times$ WFD tables). While the relative areas of the DDF compared to the WFD for LSST will be $1/400$, roughly $1\%$ of the simulated {\plasticc} data are from the DDF subset.

This light-curve tables include the following information:
\begin{itemize}
\item {\tt object\_id}: Same key as in the metadata table above, given as int32 numbers.
\item {\tt mjd}: the time in Modified Julian Date (MJD) of the observation. The MJD is a unit of time introduced by the Smithsonian Astrophysical Observatory in $1957$ to record the orbit of Sputnik. The MJD is defined to have a starting point of midnight on November $17$, $1858$. The $25^\mathrm{th}$ of September $2018$ has an MJD of $58386$. The MJD can be converted to
Unix epoch time with the formula {\tt unix\_time = (MJDœôø²40587)×86400}. The units are days, and the numbers are given as float64 numbers.
\item {\passband}: The specific LSST {\passband} integer, such that $u,g,r,i,z,y = 0,1,2,3,4,5$ in which it was viewed. These are given as int8 numbers.
\item {\tt flux}: the measured flux (brightness) in the {\passband} of observation as listed in the {\passband} column. The {\tt flux} is corrected for {\tt MWEBV}, but for large dust extinctions the uncertainty will be much larger in spite of the correction. The dust is given as a float32 number (note that the units for both {\tt flux} and {\tt flux\_err} are arbitrary).
\item {\tt flux\_err}: the uncertainty on the measurement of the {\tt flux} listed above, given as float32 number.
\item {\tt detected}: If {\tt detected}$ = 1$, the object's brightness is significantly different at the $3\sigma$ level relative to the reference template. This is given as a Boolean flag.
\end{itemize}

A few caveats about the light-curve data are as follows:
\begin{itemize}
\item \textbf{Data gaps:} Different passbands are taken at different times, sometimes many days apart.
\item \textbf{Galactic vs extragalactic:} The given redshift for objects in our own Milky Way galaxy is given as zero.
    \item \textbf{Negative Flux:} Due to statistical fluctuations (of e.g. the sky brightness) and the way the brightness is estimated, the flux may be negative for dim sources, where the true flux is close to zero. Additionally, if the pre-survey image actually contains a flux brighter than its true `zero', this can lead to a negative flux when the difference is computed.
%    \item \textbf{saturated observations} The light curves include `saturated' observations of sources, where the source is too bright to obtain a measured value. In such cases, the {\tt flux} is set to 0., and the {\tt fluxerr} is set to 10,000,000. Such an observation may not yield a value of {\tt flux}, but it indicates that the source was extremely bright rather than extremely dim at the time of observation. During operations, the LSST Project will mask out saturated observations.
%    \item \textbf{Observing Cadences} The objid string has two prefixes `DDFœôòù and `WFDœôòù.  DDF corresponds to
%œôòüDeep Drill Fieldsœôòý over a small area of sky, but with very high quality 
%light curves. WFD corresponds to œôòüwide-fast-deepœôòý over a very large 
%sky area, but with lower quality light curves.
\end{itemize}

%\clearpage
%It is possible that the distribution of properties (as found in the header table) could be different for different classes of astrophysical sources. For example, sources that are extremely dim, may only be found in our own galaxy the Milky Way, and thus their redshifts will be close to zero, and their locations are likely to be clustered around the parts of the sky where the Milky Way is densely populated. Sources that are somewhat brighter, but still too dim to be see from large cosmological distances may be found at low redshifts only. On the other hand, extremely bright sources which are visible for extremely large distances may tend to be found at higher redshifts due to larger physical volume at high redshifts. On the other hand, the signature of a particular class of astrophysical source is to be found in its light curve, which is a degraded and compressed version of its spectral evolution. 

\subsection{Obtaining the data and scoring a classification}
Participants will be required to submit a matrix of probabilistic classifications: a table with \numObjectsTest rows and {\numTotalClasses} columns (the {\numClasses} already seen in the training data and one {\tt others} class), where the sum of probabilities across all classes per row (object) is unity. Section~\ref{sec:conclusion} describes the metric that will be used to evaluate the classification probabilities.
The metric choices for {\plasticc} are described in an upcoming paper and focused on ensuring a balanced metric across classes, rather than just focusing on one source of interest.
As part of the challenge, we provide an example Jupyter notebook `starter kit' with more introductory material and another notebook to compute the metrics for the challenge\footnote{The starter kit is located at \href{https://github.com/lsstdesc/plasticc-kit}{https://github.com/lsstdesc/plasticc-kit.}}.

\subsection{Training and test data}
The training data follow the description above and have the properties and light curves of a set of {\numObjectsTraining} astronomical sources and are meant to represent the brighter objects for which obtaining expensive spectroscopy is possible. The test data represent all the data that have no spectroscopy, and is a much larger set of {\numObjectsTest} objects.
Therefore, the test data have `NULL' entries for the {\specz} column for all but a few percent of object in the test data. The {\class} column is `NULL' for all test data. Moreover, the training data properties are \textit{non-representative} of distributions of the
the test data set. The training data are mostly composed of nearby, low-redshift, brighter objects while the test data contain more distant (higher redshift) and fainter objects. Therefore, there are objects in the test data that do not have counterparts in the training data.

%%%% ----------------------------------------------------------------------
\section{Challenge participation}
\label{sec:conclusion}
{\plasticc} challenge entries are required to classify each of the sources in the header file of the test data set based on their properties and light curves. The classification is done though the assignment of probabilities $P_{ij},$ 
$P(\rm{class}\vert\rm{data}_{i}, \rm{training}~\rm{data}, \rm{knowledge}),$ 
the probability that the $i\rm{th}$ source in the test data is a member of the class $j$ based on the combination of properties and light-curve data for the object, the training data set, and any outside knowledge the participant may have acquired elsewhere, although prior knowledge of astronomy is not required to participate in {\plasticc}. To specify the entry for a single source, the participant provides the probabilities of that source belonging to each of the mutually exclusive (non-overlapping) {{\numClasses}}  classes in the training set, and of not belonging to any of the classes in the training set and therefore denoted by the {\tt others} class. High values of $P_{ij}$ for a particular class $j$ and object $i$ indicate that the participant believes that the $i$-th source is likely to be a member of the $j$th class. 
As true probabilities, the quantities $P_{ij}$ must satisfy the following criteria:
\begin{eqnarray*}
0 \leq P_{ij} \leq 1\qquad \forall i, j \nonumber \\
\sum_{j} P_{ij} = 1 \qquad \forall i
\nonumber
\end{eqnarray*}
For a {\plasticc} entry to be valid, it will have to include probabilities for each class and astronomical source, i.e. an entry cannot leave out probabilities on any source, or class. To win the challenge, all entries should be valid and minimize the {\plasticc} metric score. The metric used for the challenge is a weighted log-loss metric 
\begin{eqnarray}
  \label{eq:logloss}
  L&=& -\frac{\sum_{j=1}^{M}w_j \cdot \sum_{i=1}^N\frac{1}{N_j}\tau_{i, j} \ln(P_{ij})}{\sum_{j=1}^M w_j},
\end{eqnarray}
where $\tau_{i,j} = 1$ if the $i$th object comes from the $j$th class and $0$ otherwise, and $N_j$ is the number of objects in any given class $j,$ and $w_j$ are individual weights per class which reflect relative contribution to the overall metric (depending on e.g. how desireable it is to have objects in class $j$ classified correctly). These $w_j$ are hidden to the participants of the challenge. The averaging over $N_j$ reflects that {\plasticc} is designed to reward those who classify all objects well on average, rather than focusing on one particular class. This is discussed more in an accompanying paper (see Malz, Hlo\v{z}ek \textit{et al.} 2018 which is released with this note).

If a participant uses a classifier that decides that an object is of a particular class rather than one that provides probabilities, that participant has to use their own prescription to define probabilities from deterministic classifications. For example, a source classified as the first class may be given a probability of $1.0$ and all other classes probabilities of $0.0$, or the first class may be assigned $0.9$ and all other classes may be given uniform probabilities so that the probabilities sum to unity. 

For example, if the challenge was to classify a set of $3$ objects into two classes of `star' or `galaxy' classes (and an {\tt others} class), the returned classification table would be $3\times3$ matrix:

\begin{table}[htbp!]
\begin{center}
\begin{tabular}{|c|c|c|c|}
Object ID & $P(star)$ & $P(galaxy)$ & $P(other)$ \\
\hline
$1$ & $0.6$ & $0.3$ & $0.1$\\
$2$ & $0.3$ & $0.3$ & $0.4$\\
$3$ & $0.55$ & $0.4$ & $0.05$\\
\end{tabular}
\caption{An example classification table for a challenge to classify $3$ objects into $3$ classes. Note that the row probabilities should be normalized to unity.}
\end{center}
\end{table}

The {\plasticc} data, competition and rules for entry are described on the \href{https://www.kaggle.com/c/PLAsTiCC-2018/}{Kaggle web page for \plasticc}. The competition will feature three metric-based (`general') winners and additional science-focused prizes, and we encourage those interested to enter regardless of their experience level: previous experience in the field of astronomy is not required for participation.
%The {\plasticc} team involved in validating the data will not be able to participate in the challenge directly, and will only publish classifications on the data once the challenge has completed. 
%Some {\plasticc} team members involved in defining metrics will participate in the challenge, but they have not seen any {\plasticc} information about the models or the data.

%% file: plasticc_acknowledgements.tex
The {\plasticc} data set generation relied on numerous members of the astronomical community to provide models of astronomical transients and variables. These models will be described in a paper to be published after the challenge is complete. While we cannot thank them by name (as this could identify the models included in the challenge), we acknowledge their contributions anonymously. We thank Seth Digel, Ranpal Gill, Alex Kim and Michael Wood-Vasey for comments on this note. This work was supported by an LSST Corporation Enabling Science grant. The financial assistance of the National Research Foundation (NRF) towards this research is hereby acknowledged. Opinions expressed and conclusions arrived at, are those of the authors and are not necessarily to be attributed to the NRF. UK authors acknowledge funding from the Scieence and Technology Funding Council. Canadian co-authors acknowledge support from the Natural Sciences and Engineering Research Council of Canada. The Dunlap Institute is funded through an endowment established by the David Dunlap family and the University of Toronto. The authors at the University of Toronto acknowledge that the land on which the University of Toronto is built is the traditional territory of the Haudenosaunee, and most recently, the territory of the Mississaugas of the New Credit First Nation. They are grateful to have the opportunity to work in the community, on this territory.
%This work was completed in part with resources provided by the University of Chicago Research Computing Center. 

%% file: standard.tex
The DESC acknowledges ongoing support from the Institut National de Physique Nucl\'eaire et de Physique des Particules in France; the Science \& Technology Facilities Council in the United Kingdom; and the Department of Energy, the National Science Foundation, and the LSST Corporation in the United States.  DESC uses resources of the IN2P3 Computing Center (CC-IN2P3--Lyon/Villeurbanne - France) funded by the Centre National de la Recherche Scientifique; the National Energy Research Scientific Computing Center, a DOE Office of Science User Facility supported by the Office of Science of the U.S.\ Department of Energy under Contract No.\ DE-AC02-05CH11231; STFC DiRAC HPC Facilities, funded by UK BIS National E-infrastructure capital grants; and the UK particle physics grid, supported by the GridPP Collaboration.  This work was performed in part under DOE Contract DE-AC02-76SF00515 and with DOE award DE-DE-SC0011636.

%% file: contributions.tex
We note that while Tarek Allam, Michelle Lochner, Alex Malz, Hiranya Peiris and Christian Setzer were part of the {\plasticc} team focusing on metric selection and competition design, they have not had a role in producing, seeing, nor validating the data, and as such are allowed to compete in the challenge fully. The remainder of the team have been involved in data validation and so may participate in, but not win, the Kaggle challenge.